\documentclass[pdflatex,sn-basic,Numbered]{sn-jnl}

\usepackage{graphicx}%
\usepackage{multirow}%
\usepackage{amsmath,amssymb,amsfonts}%
\usepackage{amsthm}%
\usepackage{mathrsfs}%
\usepackage[title]{appendix}%
\usepackage[dvipsnames]{xcolor}%
\usepackage{textcomp}%
\usepackage{manyfoot}%
\usepackage{booktabs}%
\usepackage{algorithm}%
\usepackage{algorithmicx}%
\usepackage{algpseudocode}%
\usepackage{listings}%
\usepackage{cleveref}%
\usepackage{lineno}%

\usepackage{booktabs}
\usepackage{tabularx}

\usepackage{tikz}
\usepackage{tikzducks}

\usetikzlibrary{shapes.arrows, fadings}
\usetikzlibrary{arrows}
\usetikzlibrary{3d,calc}

\usepackage{hyperref}
\definecolor{kgreen}{rgb}{0.2,0.55,0.3}
\definecolor{sgreen}{rgb}{.31,.56,.38}

\theoremstyle{thmstyleone}

\theoremstyle{thmstyletwo}%

\theoremstyle{thmstylethree}%

\begin{document}

\title[Centering Ecological Goals in Automated Identification of Individual Animals]{\vspace{-2cm}Centering Ecological Goals in Automated Identification of Individual Animals}

\author*[1,2]{\fnm{Lukas} \sur{Picek}}\email{lukaspicek@gmail.com}

\author[2]{\fnm{Timm} \sur{Haucke}}

\author[1]{\fnm{Lukáš} \sur{Adam}}
\author[3]{\fnm{Ekaterina} \sur{Nepovinnykh}}
\author[4]{\fnm{Lasha} \sur{Otarashvili}}
\author[5]{\fnm{Kostas} \sur{Papafitsoros}}
\author[6]{\fnm{Tanya} \sur{Berger-Wolf}}
\author[7]{\fnm{Michael B.} \sur{Brown}}
\author[8]{\fnm{Tilo} \sur{Burghardt}}
\author[9]{\fnm{Vojtech} \sur{Cermak}}
\author[10]{\fnm{Daniela} \sur{Hedwig}}
\author[11]{\fnm{Justin} \sur{Kitzes}}
\author[11]{\fnm{Sam} \sur{Lapp}}
\author[12]{\fnm{Subhransu} \sur{Maji}}
\author[13]{\fnm{Dan} \sur{Rubenstein}}
\author[14]{\fnm{Arjun} \sur{Subramonian}}
\author[15]{\fnm{Charles} \sur{Stewart}}
\author[16]{\fnm{Silvia} \sur{Zuffi}}
\author[2]{\fnm{Sara} \sur{Beery}}

\affil[1]{\orgname{University of West Bohemia in Pilsen}, \orgaddress{\city{Pilsen}, \country{Czechia}}}

\affil[2]{\orgname{Massachusetts Institute of Technology}, \orgaddress{\city{Cambridge}, \state{Massachusetts}, \country{USA}}}

\affil[3]{\orgname{LUT University}, \orgaddress{\city{Lappeenranta}, \country{Finland}}}

\affil[4]{\orgname{Conservation X Labs}, \orgaddress{\country{USA}}}

\affil[5]{\orgname{Queen Mary University of London}, \orgaddress{\city{London}, \country{UK}}}

\affil[6]{\orgname{The Ohio State University}, \orgaddress{\city{Columbus}, \state{Ohio}, \country{USA}}}

\affil[7]{\orgname{Giraffe Conservation Foundation}, \orgaddress{\city{Windhoek}, \country{Namibia}}}

\affil[8]{\orgname{University of Bristol}, \orgaddress{\city{Bristol}, \country{UK}}}

\affil[9]{\orgname{Czech Technical University in Prague}, \orgaddress{\city{Prague}, \country{Czechia}}}

\affil[10]{\orgname{Cornell Lab of Ornithology, Cornell University}, \orgaddress{\city{Ithaca}, \state{New York}, \country{USA}}}

\affil[11]{\orgname{University of Pittsburgh}, \orgaddress{\city{Pittsburgh}, \state{Pennsylvania}, \country{USA}}}

\affil[12]{\orgname{University of Massachusetts Amherst}, \orgaddress{\city{Amherst}, \state{Massachusetts}, \country{USA}}}

\affil[13]{\orgname{Princeton University}, \orgaddress{\city{Princeton}, \state{New Jersey}, \country{USA}}}

\affil[14]{\orgname{University of California, Los Angeles}, \orgaddress{\city{Los Angeles}, \state{California}, \country{USA}}}

\affil[15]{\orgname{Rensselaer Polytechnic Institute}, \orgaddress{\city{Troy}, \state{New York}, \country{USA}}}

\affil[16]{\orgname{CNR}, \orgaddress{\city{Milan}, \country{Italy}}}


% \linenumbers

\abstract{
Recognizing individual animals over time is central to many ecological and conservation questions, including estimating abundance, survival, movement, and social structure. 
Recent advances in automated identification from images and even acoustic data suggest that this process could be greatly accelerated,  yet their promise has not translated well into ecological practice.
We argue that the main barrier is not the performance of the automated methods themselves, but a mismatch between how those methods are typically developed and evaluated, and how ecological data is actually collected, processed, reviewed, and used. 
Future progress, therefore, will depend less on algorithmic gains alone than on recognizing that the usefulness of automated identification is grounded in ecological context: it depends on what question is being asked, what data are available, and what kinds of mistakes matter.
Only by centering these questions can we move toward automated identification of individuals that is not only accurate but also ecologically useful, transparent, and trustworthy.
}
\maketitle  
\section{Introduction}
\label{sec:intro}

Repeated observations of identifiable individual animals are of great value in ecology and conservation~\cite{clutton2010individuals,vidal2021perspectives,brown_male-biased_2020}. They support studies of, for example, behavior, health, movement, migration, communication, kin structure, population dynamics, or drivers of population change, while also informing conservation assessment and decision-making, for example, through population monitoring and IUCN Red List evaluations~\cite{choo2020best,rubenstein2016equus,muller_giraffa_2018,balazs1999factors}.
To generate these repeated observations, individual identification surveys traditionally require capturing and marking animals (\textit{mark--recapture}) so that they can be recognized on future sampling occasions, e.g., through application of metal bands, PIT tags, collars, or even live tissue tags \cite{lettink2003introduction,lindberg2012review}. 
Although widely used, capturing and handling the animals raises ethical concerns and can affect behavior or survival \cite{bodey2018phylogenetically}. 
These methods also require substantial field effort and logistical support, which can limit sample size, spatial coverage, and study duration~\cite{tucker2010nest}. Even with this investment, markers may be lost, and recapture or resighting rates may remain low, reducing the completeness and reliability of encounter histories~\cite{pfaller2019genetic}.

In contrast, non-invasive identification relies on unique and persistent natural features, which we refer to as \textit{signatures}, that allow individual recognition without capturing and manually marking the animals \cite{kuhl2013animal,karczmarski2022individual,vidal2021perspectives}, and thus can scale more easily across sites and seasons. 
Such signatures can be based on visual aspects (e.g., skin, coat, scale, fin patterns, scars, spots), motion (e.g., gaits), physical signs (e.g., DNA, hair traps, scat, footprints), or even vocal traits (e.g., call characteristics, unique songs or whistles)~\cite{lahiri2011biometric,schofield2008investigating,andreotti2016integrated,Knight2024AIID,foster_giraffe_1966}. 

While this is more clearly feasible for some species, such as those with spots or stripes, it remains unclear whether non-invasive individual identification is possible for the vast majority of taxa, especially beyond the best-studied visual cases. In many species, potentially useful signatures may instead come from other modalities, such as vocalizations, movement, footprints, chemical cues, or genetic traces, but their value for reliable long-term identification remains poorly understood. Even for identifiable species, natural signatures may change over time and be difficult to observe because of pose, molt, injuries, or data quality issues, all of which can lead to misidentification. Thus, reliable individual identification often requires trained expertise and extensive time and effort~\cite{carpentier2016stability}.
 
In recent years, as imaging and bioacoustic technology have improved \cite{hoefer2023passive,sugai2019terrestrial}, non-invasive long-term wildlife studies have scaled up dramatically. Wildlife monitoring now often relies on sensor-based approaches using camera traps, drones \cite{dujon2021machine}, passive acoustic recorders, images and recordings from social media \cite{toivonen2019social}, citizen science projects, and widely available cameras and smartphones.
These approaches generate large numbers of records across space and time, which significantly improve our ability to document when and where animals occur.
Many of these studies focus on species-level questions, such as occupancy, activity, and community composition, rather than on tracking and studying unique individuals. Only a small fraction of these studies focus on individual identification.
As with non-invasive approaches, a large share of today's data is still only used to characterize animal presence or activity rather than individual life histories. 

If we could identify individuals across this wealth of collected data, we could answer much richer demographic and behavioral questions that have traditionally relied on invasive mark--recapture methods. 
There have been significant advancements in automated AI-based individual animal identification in the last few years~\cite{schneider2019past,ravoor2020deep}. Many of the developed systems promise automatic and reliable recognition of individual animals from photos, videos, and even sound recordings, and many papers report very high identification performance.
But these systems are still rarely used as a routine part of ecological monitoring.

\begin{figure}[h]
    \centering
    \includegraphics[width=\linewidth]{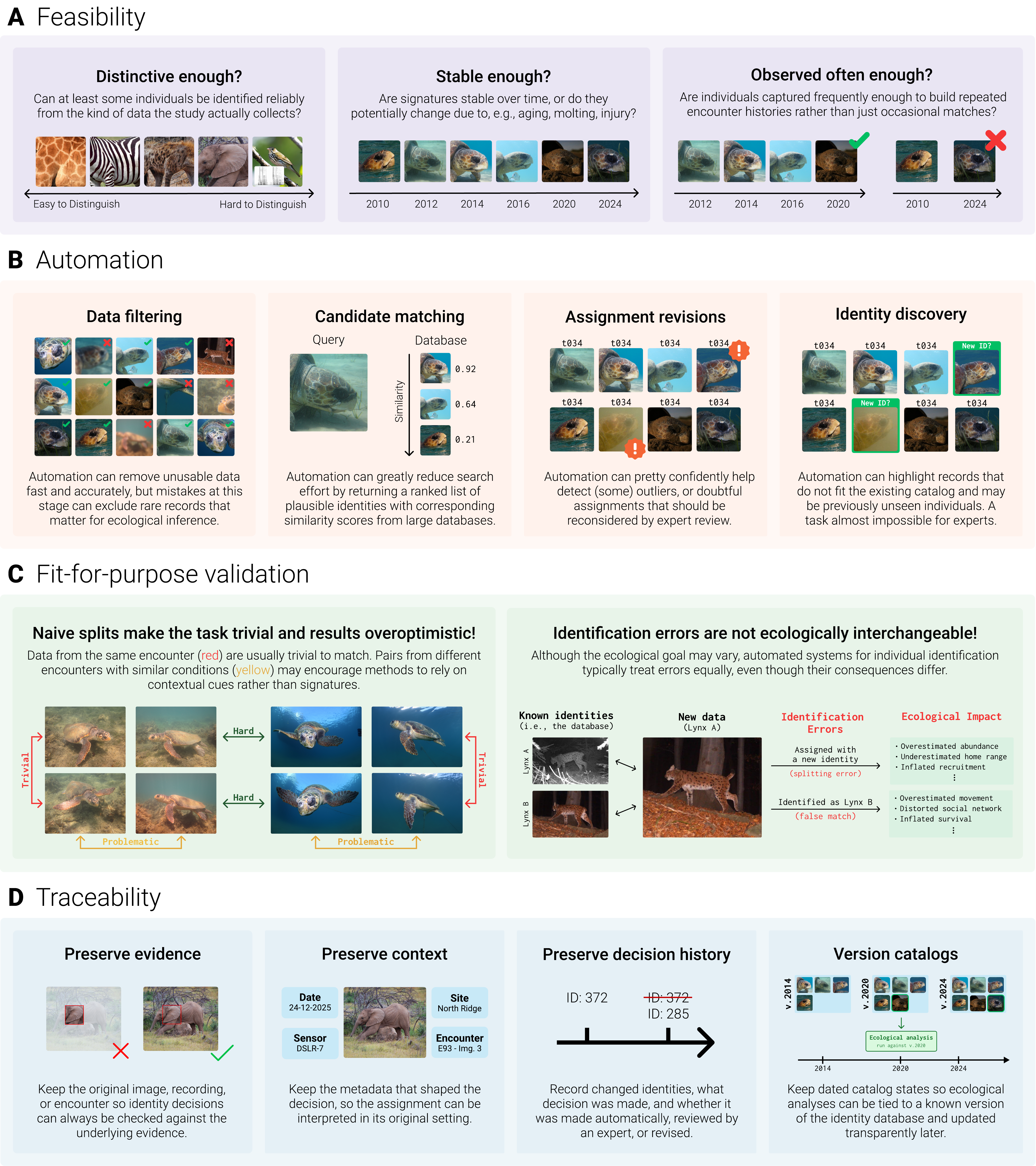}
    \caption{
    \textbf{Practical considerations for ecologically useful automated individual identification.} Before using any method, adopters should at least ask themselves four connected questions. \textbf{A}: \textit{Is identification feasible for the species, timescale, and data at hand?} \textbf{B}: \textit{Can automation save time, and where should expert judgment be focused?} \textbf{C}: \textit{Which identification errors would most affect our ecological question?} \textbf{D}: \textit{What needs to be recorded so identity assignments stay transparent, revisable, and trustworthy over time?}
    }
    \label{fig:hero}
\end{figure}

\bigskip
\noindent \textbf{In this perspective, we argue that making automated identification of individual animals ecologically useful depends less on algorithmic novelty alone than on centering ecological goals in how methods are designed, evaluated, deployed, and integrated into study design. Rather than treating identification as a purely technical problem, we frame it around four connected practical questions (Figure \ref{fig:hero}): (i) is identification feasible, (ii) where does automation help, (iii) which errors matter most for the ecological goal, and (iv) what should be recorded so decisions are transparent, revisable, and trustworthy over time.}

\subsection*{The Missing Links: Why Automated Individual Identification Has Not Translated Well to Ecological Practice}

What is driving the gap between methods development and practical deployment? We argue that it is primarily related to misaligned objectives. Below, we highlight three main gaps that limit broader adoption.

\begin{enumerate}
    \item \textbf{Mismatch between ecological reality and how methods for automation are developed and evaluated.}
    In ecological studies, populations are rarely closed. Individuals are born, die, immigrate, and emigrate, and observations are collected over long periods of time, across sites, seasons, sensors, and changing field conditions. Encounters are often sparse, unevenly distributed across individuals, and difficult to compare directly across years or locations. In this setting, the same identification error can have very different consequences depending on the ecological question, such as abundance estimation, survival, movement, or social structure.
    Yet many automated methods are still developed and evaluated under much simpler conditions: a fixed list of known individuals, relatively homogeneous data, and test settings that do not reflect temporal, spatial, or sensor-related change~\cite{adam2025wildlifereid,vcermak2024wildlifedatasets}. These assumptions can make performance appear stronger than it will be in practice and can obscure whether a method is actually suitable for ecological use. To be useful in the field, automated identification systems should therefore be developed and evaluated with ecological deployment in mind: whether they can handle previously unseen individuals, remain robust to change over time and across conditions, and are assessed in terms of the kinds of errors that matter most for the intended ecological application.
    \vspace{2mm}

    \item \textbf{Mismatch between ecological workflows and the push for full automation.}
    In ecological practice, identification is rarely a fully automated decision based on a single image or recording. Instead, experts often review encounters that contain multiple observations of the same animal and interpret them in light of contextual knowledge, such as group composition, site familiarity, expected movements, home ranges, or previous encounter history. Identification is therefore often a process of weighing evidence rather than assigning a label in isolation.
    By contrast, automated methods are often framed and evaluated as if their main value lies in replacing expert judgment with a single automatic output. This overlooks the fact that current systems usually do not make use of much of the context that experts rely on. As a result, small changes in pose, occlusion, or lighting may cause automated systems to fail even when an experienced observer would still make a reliable decision~\cite{dunbar2021hotspotter}. At the same time, automated systems may detect subtle visual differences that are difficult for humans to notice~\cite{tuia2022perspectives,adam2025exploiting}. Their greatest value may therefore lie not in replacing experts, but in supporting them: for example, by retrieving a short list of candidate individuals, flagging uncertain cases, or highlighting inconsistencies across time and space~\cite{inkpen2023advancing}. Designing methods, interfaces, and evaluation frameworks around this \textit{expert-in-the-loop} reality is essential for linking automated identification to ecological practice~\cite{perez2024human}.
    \vspace{2mm}

    \item \textbf{Limited attention to resource constraints and practical costs.}
    Ecological studies operate under broad resource constraints, including limited time, funding, personnel, expertise, annotation capacity, computing infrastructure, and long-term data-management support~\cite{stewart2021animal}. In this setting, the key question is often not whether one method performs slightly better than another, but whether a method is practical, affordable, and sustainable for a particular study.
    This makes adoption of automated identification a practical as well as a technical question. Before choosing a method, researchers need to understand what it will require from their own data, team, and workflow.
    Specifically, they need to know how much expert review will still be needed, how much labeled data must be assembled, what hardware or cloud resources are necessary, what software and technical skills are required, and what ongoing effort is needed to maintain catalogs, update models, and document decisions over time. Method choice is therefore also a question of resource allocation: when is it most useful to invest in new annotations, expert review, improved sampling design, computational infrastructure, or downstream ecological analysis? Greater attention to these practical costs and trade-offs would make automated identification easier to evaluate, adopt, and trust in ecological applications.
\end{enumerate}

\medskip
\noindent\textbf{The path forward.}
This paper brings together perspectives from ecology, conservation biology, and machine learning to help practitioners better understand the complexities and practicalities of AI tool use on their own data and field conditions, and to integrate automation into expert-reviewed workflows and downstream population analyses.
We argue for a shift toward \emph{fit-for-purpose evaluation}: assessing methods in terms of the ecological question, study design, and tolerance for different identification mistakes, rather than relying mainly on performance measured on \emph{non-standardized data collections}.
By linking ecological goals, sampling design, and \textit{expert-in-the-loop} workflows to the strengths and limitations of current tools, we aim to lower barriers to adoption and increase the value of automation, for example, in population and demographic studies and, ultimately, for biodiversity conservation.
\section{Understanding the Current Status Quo}

To ensure an interdisciplinary audience is on the same footing, we start with a brief review of the types of ecological studies that are informed by the identification of individual animals, provide a brief historical context on the development of both manual and automated approaches for this challenge, and characterize the current framing of the problem in the machine learning community. Since these communities often use different terms that refer to the same or similar concepts (which further reflects the mismatch between method development and practical usage), we include a brief glossary (Table~\ref{tab:glossary}) that lists and translates the main terms we rely on across both disciplines.

\label{sec:positioning}

\subsection{How Identification of Individuals is Used in Ecology}\label{sec:ecological_uses}

Individual identification is rarely the final goal of an ecological study. In many long-term projects, these individual encounter histories are the key link between field observations, statistical models, and management decisions. Once individual animals are recognized, their encounters over time and space become the raw material for several questions about their ecology, biology, and conservation \cite{lindberg2012review}. 
Examples of such questions include:

\begin{itemize}
    \item \textbf{Abundance and population trends.} Distinguishing individuals within a population allows estimating population size and tracking how it changes over time by observing which individuals appear, disappear, or newly enter the study area. For many threatened or protected species, such individual-based estimates are required for assessing progress toward recovery targets and/or assessing the impact of conservation actions \cite{WilliamsNicholsConroy2002,arzoumanian2005astronomical, hudgins2023brighter,brown2019all}.

    \vspace{1mm}
    \item \textbf{Birth, death, and demography.} Following identified individuals across years supports estimates of survival, age structure, recruitment, and life expectancy, and allows comparison of demographic rates across sex, age, or status classes \citep{WilliamsNicholsConroy2002,Schofield_2020,brown2025assessing}.

    \vspace{1mm}
    \item \textbf{Movement and connectivity.} Linking the same individual across space and time provides non-invasive information on territories, home-range use, dispersal, migration routes, and connectivity among sites or subpopulations \cite{brown_male-biased_2020,ovaskainen2008bayesian,montagna2023combined}.
    
    \vspace{1mm}
    \item \textbf{Behavior and social structure.} Knowing which individual is doing what, and which individuals are seen together, enables studies of behavior (e.g., foraging, mating, parental care, aggression) and social organization, including dominance hierarchies, kinship, group membership, and how social networks change over time \citep{schofield2022more,bond2021proximity}.
    
    \vspace{1mm}
    \item \textbf{Health, disease, and disturbance.} Repeated observations of the same individual, especially when physical condition or symptoms are visible, allow tracking disease progression, injury histories, recoveries, and the effects of human disturbance or tourism \cite{papafitsoros2021social,hancock2023using,fong2026long}.
    
    \vspace{1mm}
    \item \textbf{Variability in habitat use and preferences.} While habitat preferences are often estimated at the population level, linking locations to individuals reveals within-population variation in habitat use, plasticity, competition, and behavior~\cite{dujon2018complex}.
    
    \vspace{1mm}
    \item \textbf{Human--wildlife conflict and management.} In systems where specific individuals cause conflicts with people (e.g., crop raiding, livestock predation, or property damage), being able to identify the responsible animals enables targeted management rather than broad population-level control~\cite{swan2017ecology}.
    
    \vspace{1mm}
    \item \textbf{Public awareness.} Long-term identification of focal individuals allows individual stories to be used in outreach and public awareness campaigns, where sharing the lives of recognizable animals helps engage the public and build support for conservation \citep{jaric2024flagship}.
\end{itemize}

Despite these applications sharing the same basic requirement (e.g., recognizing individuals), they differ in how identification errors propagate into ecological inference. At a high level, two types of identification error are especially important.
First, merging encounters from different animals into a single identity (a \emph{false match})
can negatively bias abundance at the population level, while inflating survival estimates and exaggerating movement distances for that specific single identity.
Second, assigning encounters from the same animal to multiple identities (a \emph{missed link}) can positively bias population abundance and recruitment, deflate survival, and make its home range and dispersal distances appear smaller.
% \medskip
% \noindent\textbf{Understanding what matters:} 
This asymmetric impact of errors is not well-quantified across populations and ecological tasks, but it dictates which methods might be suitable for a particular ecological goal. \textbf{In this work, we therefore center the discussion on ecological implications when framing how tools and methods for automation of individual identification should be designed and evaluated for ecological use.}

\subsection{How Individual Identification Has Been Framed Over the Years}
\label{sec:positioning-prior}

\noindent Building on decades of manual photo-id in field studies and on early computer-vision-assisted matching systems developed in the 1990s~\cite{hiby1990computer}, a later wave of descriptor-based tools such as Wild-ID~\cite{bolger2012computer}, I3S~\cite{speed2007spot}, and HotSpotter~\cite{crall2013hotspotter} provided tools for automated matching of individuals with distinctive marks, like stripes, spots, and contours, using hand-crafted image features. These systems popularized the idea of ranked \emph{retrieval} as the goal of individual identification systems (i.e., returning a short list of candidate matches for human confirmation) and proved effective for strongly marked taxa with relatively stable views~\cite{arzoumanian2005astronomical,town2013m,dunbar2021hotspotter}. These early methods had well-documented limitations, including sensitivity to viewpoint, illumination, occlusion, and changes in markings over time~\cite{carpentier2016stability,halloran2015applying}.

Over time, these studies, tools, and platforms have been producing large archives of raw data (particularly images), encounter histories, and match decisions. Much of this information was originally stored in project-specific databases, spreadsheets, or software, with different formats, naming schemes, and levels of metadata. Parts of these archives have been shared through public datasets and community platforms that now cover many species and sensing conditions and support uploading, curating, and matching images and recordings at scale~\cite{vcermak2024wildlifedatasets,berger2017wildbook,cheeseman2017happywhale}. Some datasets follow individuals over decades and include timestamps and site or camera metadata~\cite{adam2024seaturtleid2022,picek2026czechlynx}, whereas others require substantial curation~\cite{vcermak2024wildlifedatasets}.

With the growth of labeled data, research on automated animal identification has shifted from hand-crafted matching toward deep learning. These methods are usually trained in one of two ways.
Some are trained to assign each sighting one of a fixed set of known individuals (i.e., \emph{classification}), whereas others learn a feature space in which sightings of the same individual are closer together, and sightings of different individuals are farther apart (i.e., \emph{metric learning}) \cite{vcermak2024wildlifedatasets, otarashvili2024multispecies}.
In practice, however, both are often used in a similar way: the system produces a vectorized representation of new data and returns a ranked list of candidate identities for human review. 
For broader overviews, see \cite{ravoor2020deep} and \cite{schneider2019past}.

\subsection{What Makes Automated Identification of Individuals Difficult}

Even with modern methods, automated identification of individuals in wild animal populations is a hard problem. 
Unlike in idealized biometric settings, the available signatures are often not fully unique, persistent, measurable, or consistently comparable~\cite{jain2005biometric} due to:

\begin{itemize}
    \item \textbf{Open populations.} New individuals are born or immigrate into the study area, and others die or emigrate. Any system must be able to recognize when an observation belongs to an individual that has never been seen before, rather than forcing every record into an existing identity. This extends more traditional biometric recognition settings, e.g., face and fingerprint verification~(\textit{1-to-1} \textit{matching}) and face recognition of known individuals~(\textit{1-to-n matching}) into the realm of online, with an unknown number of individuals to recognize ~(\textit{1-to-? matching})~\cite{picek2026czechlynx,rosenberg2026individual}.

    \vspace{1mm}
    \item \textbf{Changes in signature.} Individuals do not look or sound the same across time and space. In images, 
    appearance can change due to growth, molting, injuries, mud, snow, or epibionts on the body. 
    Acoustic signals can change due to open-ended learning, morphological changes, or physiological changes such as breeding readiness \cite{rivera2012tuning,recalde2025demographic}.% % 
    
    \vspace{1mm}
    \item \textbf{Identifying animals in the wild.} 
    For many species, individual identification is only possible when a particular feature is recordable. \textit{However, animals do not pose for cameras.} They are often occluded by vegetation or terrain, moving, turned away, showing body parts not useful for identification, or communicating with unidentifiable calls with background noise and even including other individuals and sounds~\cite{grolleau2026moo,rosenberg2026individual}.
    
    \vspace{1mm}
    \item \textbf{Heterogeneity in observation.} Often, a few individuals in a population are photographed or recorded many times, while others are seen only once or twice, sometimes years apart. These long-gap re-observations can be the most difficult to match, but matter most for estimating survival or long-distance movement~\cite{picek2024animal}.

    \vspace{1mm}
    \item \textbf{Heterogeneity in identifiability.} As in classic \textit{mark--recapture} and photo-based identification work, some animals are simply easier to detect and recognize than others (e.g., an individual with a unique signature or injury) \cite{palmer2015contextual}. This can compound with the \textit{heterogeneity of observation} and \textit{difficulty of identifying animals in the wild}, making an elusive or rarely seen, less visually distinctive member of a population easier to mistake~\cite{halloran2015applying}.
    
    \vspace{1mm}
    \item \textbf{Shifts in site, season, and sensors.} 
    Field data span across locations, habitats, seasons, and devices. Background and soundscapes, vegetation, weather conditions, and sensor-specific characteristics all change over time. These shifts can confound AI methods, leading to degradation of performance over time for the same population~\cite{beery_recognition_2018,picek2024animal}.
    
\end{itemize}

Together, these factors mean that fully automated systems will inevitably make mistakes, especially on the rare and difficult cases that are often most informative for ecology. The practical question is therefore not how to eliminate errors, but how to design systems and workflows so that errors are made in the least damaging cases, are detectable and correctable, and are accounted for when drawing ecological conclusions. 

\subsection{What Current Evaluation Rewards, and What it Overlooks} Much of the work on automated individual identification in the machine learning community is driven by improving scores on non-standardized datasets~\cite{orr2024ai,vcermak2024wildlifedatasets}. These benchmarks are important: they make it easier to share code, compare methods fairly, and track progress over time~\cite{adam2025wildlifereid}. However, the way they are currently designed often does not realistically reflect how ecologists want to use automated methods (i.e., for the questions in Section~\ref{sec:ecological_uses}).

\medskip \noindent \textbf{How data is selected for training and evaluation.} 
In machine learning, datasets are typically divided into separate subsets for \emph{training}, \emph{validation}, and \emph{testing}. The \textit{training} set is used to fit the model, the \textit{validation} set is used to tune decisions during development, and the \textit{test} set is held back for final evaluation. 
This is fine in principle, as it assesses whether a method performs well on data it has never seen. However, the way these splits are constructed often does not match the realities of ecological use.
First, many methods are developed and evaluated under a \emph{closed-set} assumption, in which every sample in the test set is assumed to belong to one of the individuals available in the training data. In ecological studies, by contrast, individuals appear and disappear regularly over time through birth, death, immigration, and emigration.
Second, the training--validation--test splits are often created randomly. As a result, nearly identical samples from the same encounter might be used for both training and testing, inflating performance compared with what it would be in later seasons, at new sites, or on different devices.
Even when exact duplicates are avoided, samples collected close together in time or under very similar conditions may still be much easier to match than the future data the method will face in practice.

\medskip \noindent \textbf{How performance is measured on evaluation data.} 
Most benchmarks uniformly summarize performance across all samples in the \emph{test set}, often measuring only accuracy (i.e., the proportion of test samples assigned to the correct identity). As a result, often-photographed individuals have a much greater influence on the final score than those rarely seen. In contrast to the evaluation, much of the expert effort in ecological studies is spent on difficult cases, such as long-gap re-sightings, individuals whose markings have changed, or animals seen only briefly or under poor conditions. Benchmarks rarely report how well methods perform on such rare or high-consequence cases, or how much expert effort is required to review candidate matches, reject poor matches, or flag likely new individuals. Yet these are often the factors that determine whether a method can support reliable estimates of abundance, survival, movement, or social structure. Finally, identification errors are not interchangeable: merging two individuals and splitting one individual into several identities can bias ecological inference in different directions, but this distinction is rarely reflected in traditional performance metrics.

\section{Practical Considerations Before Deployment}
\label{sec:considerations}

The main challenge in automated individual identification is not simply methodological performance. It is whether a workflow that relies on automation can support an ecological question under real field conditions. 
We identify four key practical considerations that we find are often missed before diving headfirst into developing automated methods for identification of individuals (see Fig.~\ref{fig:hero}), and outline a workflow for assessing each.

\subsection{Feasibility of Individual ID for a Particular Species and Data} 

For some species, individuals may not differ much, signatures may change too quickly, or the available data may be too inconsistent or low quality for robust identification. This is known as irreducible error in machine learning. In this case, no automated system will recover information that is not present in the data, and in the worst case, automated systems may provide high-confidence matches that, if not verifiable, should not be trusted. The key initial focus of any project should therefore be whether the species, timescale, and sampling design provide a usable and verifiable signature for individuals \cite{schofield2008investigating}.
Before choosing a method, tool, or building an automated workflow, the following three questions should be asked:
\begin{enumerate}
    \item \textit{Can a trained observer consistently identify at least some individuals from the kinds of data the study actually collects (e.g., camera traps), rather than from a small set of ideal examples (e.g., wildlife photography)?}

    \vspace{1mm}
    \item \textit{Are signatures stable over time, or do they potentially change due to, e.g., aging, molting, injury, or seasonal variation?}
    
    \vspace{1mm}
    \item \textit{If needed for the intended ecological analysis, are signatures captured sufficiently frequently and clearly to produce repeated encounter histories, rather than just occasional successes?}
\end{enumerate}

\textbf{If the answer to these questions is mostly ``no'', then the main bottleneck is likely biological or sampling-related, not algorithmic.}
However, even if the answer is mostly ``yes'', this establishes only that individual identification may be possible in principle. It does not imply that any kind of automation will be feasible, accurate, or cost-effective in practice.
Modern automated systems often require substantial labeled data, representative variation in training, and evidence of reliable performance for the intended ecological use.

Last but not least, feasibility is often partial and study- and species-specific rather than all-or-nothing. A species may be identifiable only as adults, from one body side, during some seasons, or under certain viewing conditions. That does not rule out individual identification, but it should inform data collection protocols and/or constrain which ecological questions can be answered reliably. In such cases, the goal should be to define clearly which encounters are informative, which are not, and how those limits affect downstream inference.

\subsection{Where Automation Helps, and Where Humans Should Make the Final Decisions}
If individual identification is possible, the next question should be where automation might help and where expert judgment should remain central. In most ecological studies, automation might not be able to replace expertise completely, but it may reduce routine work: removing unidentifiable data, retrieving likely identities, and flagging cases that appear inconsistent over time, across space, or in prior encounter history.
\textbf{In any case, expert review should remain central wherever identification decisions could change the ecological result}, particularly for sensitive or threatened species.
This could include deciding whether a sighting belongs to a known individual or represents a new one, resolving ambiguous or low-quality cases, and reviewing matches that would strongly affect estimates of abundance, survival, movement, dispersal, or social structure.  These are not just technical edge cases; they are the points where identification uncertainty is most likely to become ecological bias. 
In practice, this means expert effort should not be evenly distributed across recordings. 
\textbf{Ideally, automation should handle volume, while expert effort should focus on the decisions that matter most for the ecological goal.}

\subsection{Which Mistakes Matter Most for the Ecological Question}

If individual identification is feasible and the role of automation is clear, the next question is which identification mistakes would do the most damage to the ecological analysis. In some studies, avoiding false matches should be the priority (\textit{precision--sensitive}); in others, avoiding missed links could matter more (\textit{recall--sensitive}). Merging multiple individuals into a single identity and splitting a single individual into multiple identities create distinct, often opposing, biases (Table~\ref{tab:errors}). A system that appears accurate overall may still be poorly suited to a study if it makes the wrong kind of mistakes.
In population estimation, false matches can bias abundance downward, whereas missed links can inflate apparent abundance or recruitment \cite{yoshizaki2009modeling}. For movement, dispersal, or site fidelity, missed links can be especially damaging because they break encounter histories across time or space, while false matches can create movement events that never occurred. For behavior or social structure, both error types can distort association patterns by either inventing links or erasing ones.

Studies should therefore define error priorities before choosing metrics, thresholds, or review rules. The key question is not whether a method has high average performance, but whether its performance is well-aligned with the intended ecological use. \textbf{Fit-for-purpose evaluation is therefore not optional. It is the only way to judge whether automated identification models can be safely and reliably deployed in a particular ecological study.}

\subsection{What Should be Recorded so Identities Remain Revisable and Reliable Over Time?}

Encounter histories do not emerge automatically from raw observations. 
Rather, they are assembled through a chain of data processing, candidate matching, expert review, and later correction. \textbf{Identity records should therefore be treated as revisable scientific products rather than fixed facts.} The key question is simple: if an important identity assignment is questioned later, can the study reconstruct how that decision was made and what information was available at the time?
In practice, this requires a clear decision history. Authors should be able to recover the original image, recording, or encounter; the relevant metadata, such as date, site, and sensor; the candidate identities considered; the final decision; and whether that decision was made automatically, reviewed by an expert, and/or later revised \cite{wilkinson2016fair}. This matters especially in long-term studies, where catalogs grow, marks change, and earlier assignments may need correction.

Changes should also remain visible. When identities are merged, split, or relabeled, the earlier state should not disappear without a trace. Stable identifiers, dated catalog versions, and simple change logs allow analyses to refer to a known catalog state and to be updated transparently when the identity history changes.
Any assignment that could materially affect abundance, survival, movement, or social inference should remain reviewable, ideally in perpetuity. In that sense, traceability should be part of ecological rigor \cite{brun2025enabling}. It allows encounter histories to improve over time, and it makes automated identification more trustworthy as methods, data, and projects evolve. However, we currently lack the necessary off-the-shelf systems or data management tools to support this level of reproducibility and traceability without adding a substantial burden to data collectors and managers. Achieving this standard today would usually require dedicated database infrastructure, persistent identifiers, versioned data management, clear annotation and review protocols, and people with the time and technical capacity to maintain these systems over the long term. This outlines a clear priority for technological innovation moving forward.
\section{Framing Existing Projects Through this Lens}
\label{sec:stories}

Real deployments of automated individual animal identification rarely follow an ideal plan. Field logistics, annotation effort, shifting study goals, and incomplete metadata all leave lasting traces in the data, the workflow, and the effort required to achieve the intended outcome. 
We illustrate this using two case studies. Both evolved over time as researchers responded to practical constraints, new questions, availability of new methods and tools, and growing data scales of both number of individuals and observations per individual considered.

\subsection{Grevy's Zebra: A Precision-sensitive Mark--recapture Setting}

The Great Grevy's Rally (GGR) is a biannual photographic census of Grevy's zebra in Kenya, carried out since 2016 \cite{ggr}. During each rally, citizen scientists drive prescribed routes over two survey days and photograph zebras encountered along the way. 
The survey estimates population size from the numbers of individuals seen only on Day 1, only on Day 2, and on both days, using the Lincoln–Petersen estimator \citep{lincoln-petersen}.
Because the survey window is short, the analysis assumes no recruitment, perfect survival over the two days, no migration, and equal sighting probability.
This makes the system a clear example of how identification errors can affect ecological inference, as both false matches and missed links can bias the abundance estimate.

A central practical issue is deciding which detections should enter the matching process at all. Some exclusions are obvious, such as heavily occluded animals, blurry images, etc. Others are less clear, e.g., two images may show the same zebra but from non-comparable views, or they may show the same side at angles or under occlusions that still prevent confident identification. In response, the workflow developed the notion of an \emph{identifiable annotation}: a detection that clearly shows a consistent, matchable marking region. This emerged from the practical need to avoid wasting effort on unmatchable comparisons and to reduce bias introduced by forcing poor-quality detections into the identification process. GGR initially implemented this step manually and later automated it with a machine-learning-based classifier.

Once identifiable records are selected, they must be grouped into sets of observations corresponding to unique individuals. Early rallies relied on HotSpotter~\citep{crall2013hotspotter}, whereas more recent work uses deep embeddings from MiewID \citep{otarashvili2024multispecies}, which make similarity search and clustering more computationally tractable over larger sets of data, in terms of both the number of individuals and the number of observations per individual. Even today, the workflow is far from fully automated. Instead, automation is used primarily to filter for high-quality, identifiable observations, reduce search effort over large populations, and organize candidate matches. Every identity match still receives expert review and verification, but automation makes this tractable.

\medskip
\noindent\textbf{Takeaway:} \textit{The GGR use case shows how the proposed workflow was shaped by long-learned practice, i.e., identifiability must first be defined, unhelpful detections excluded, and automation is used selectively to reduce but not replace human verification. This illustrates that ecologically useful identification of individuals is not achieved by a fixed pipeline but by an ever-evolving, adaptive workflow.}

\subsection{Eurasian lynx: A Recall-sensitive Setting for Long-term Survival and Movement}

Since 2015, a camera-trapping project has monitored Eurasian lynx (\textit{Lynx lynx}) across the Czech-Slovak-Polish borderland \citep{dula2021multi}. Unlike short-term mark--recapture surveys, the goal in this setting is to maintain encounter histories across seasons, sites, and changing field conditions so that survival, turnover, and movement can be inferred over time. Because failing to reconnect two encounters of the same individual can break an apparent survival history, erase a movement event, or obscure a territory shift, this is a recall-sensitive setting, even though false matches still need to be controlled. That, in turn, makes the main practical challenge not fully automatic identification, but recovering plausible re-encounters from a growing catalog in a way that experts can trust.

The current workflow is built around expert review, which makes automation useful mainly as support rather than a replacement. Earlier steps, such as filtering empty images and sorting species, are already much easier to automate. The harder problem is adding automation at the individual-identification stage, where methods such as WildFusion \cite{cermak2024wildfusion}, MegaDescriptor \cite{vcermak2024wildlifedatasets}, and ranking-based retrieval are being explored as support tools to search the identity catalog and return plausible candidate individuals for expert review.

Here, realistic evaluation matters because it is what will make automation credible to experts. The question is not whether a model performs well on average, but whether it performs well on the cases that matter in this workflow: long-gap re-encounters, cross-site matches, changing seasonal conditions, and incomplete or uncertain labels. That is why the CzechLynx dataset \cite{picek2026czechlynx} was built not with the goal of creating a clean benchmark, but to test whether automation can work under the same messy conditions that experts already deal with in practice. In this setting, future adoption of automation will depend on showing that it can reduce search effort without weakening trust in long-term encounter histories.

\medskip
\noindent\textbf{Takeaway:} \textit{When automation is not yet well established in a workflow, fit-for-purpose evaluation should show not only that the right re-encounters can be recovered, but also that expert time can be reduced without weakening trust in long-term encounter histories.}
\section{Perspectives for Ecologically Useful Identification of Individual Animals}
\label{sec:conclusion}

\textbf{Our main message is pragmatic and hopeful:} automated identification of individual animals does not need to be perfect to be scientifically useful.
What matters, across species and ecological studies. is not whether a method achieves high performance, but whether it is fit for the ecological purpose it is meant to serve. That depends on the question being asked, the kind of data available, the realities of field collection, and the consequences of the mistakes the system is likely to make. Because those consequences differ across ecological applications, automated systems should be judged less by average accuracy alone than by whether they support reliable ecological inference and reduce effort in practice.
Before introducing any level of automation, we should therefore ask: 
(i) whether individual identification is feasible for the species and data at hand, 
(ii) where automation would add the most value, 
(iii) which errors matter most for the intended use, and 
(iv) what information must be preserved in encounter histories and identity catalogs to ensure reproducibility.

Taken together, these points suggest that the next advance in automating the identification of individual animals will come not only from better algorithms but also from better alignment with ecological goals. In most ecological scenarios, the most effective systems will not replace experts, but will focus expert effort where judgment is most needed and where errors have the greatest ecological consequences. By matching tools to the context of ecological studies, retaining expert input where it matters most, and treating identity data as shareable, versioned artifacts, we can transform raw nature observations into robust ecological insight and strong evidence for biodiversity conservation.

\medskip
\noindent\textbf{What can we do now?}
Existing methods are already scientifically useful in some common ecological settings. In repeated photographic censuses with clear natural marks, short sampling windows, and well-controlled effort, current methods can support estimation even when some links are missed, provided that large counting errors remain limited. In long-term studies of open populations, automated tools can also reduce expert workload substantially by narrowing candidate lists, flagging likely new individuals, and helping maintain encounter histories, while leaving higher-impact decisions to expert review. Across both types of settings, workflow design decisions often matter as much as the particular identification model used. These include how encounters are defined, when a new identity is created, how merges and splits are recorded, how unusable observations are handled, and whether evaluation reflects realistic deployment conditions. With these pieces in place, current methods can support robust inference about, for example, abundance, survival, and movement.

\medskip
\noindent\textbf{What are the next steps?}
The next step is not to wait for a universal recipe for automated individual identification, because no single checklist will determine in advance when automation is appropriate, how much expert review is needed, which errors are tolerable, or how identities should be updated over time. \textit{These decisions require critical thinking and expert knowledge} as they usually depend on the species, the data, the study design, and the ecological question. Instead, the field should focus on building workflows and tools that make assumptions explicit, preserve uncertainty, and support revision as evidence accumulates. This will require flexible, easy-to-use tools and systems that help maintain consistency, transparency, and traceability without substantially increasing the burden on already stretched-thin data collectors.

Just as importantly, the field should invest in stronger interdisciplinary capacity and communities around individual animal identification. Progress will depend on computer scientists who understand ecological questions and the consequences of model errors, and on ecologists’ deepening understanding of identification methods and their limitations. Building this capacity will require shared, targeted training, closer integration of ecological and technical expertise in study design, and collectively defined community standards that make assumptions, uncertainty, and downstream use easier to communicate across disciplines.

\appendix

\newpage
\section{Influence of Identification Errors on Abundance Estimates}

A central argument of the main paper is that identification errors are not all equal: their consequences depend on the ecological question. For readers less familiar with ecological inference, we illustrate this point here using one of the simplest and most widely used abundance estimators, the Lincoln--Petersen estimator \cite{lincoln-petersen}. The goal is not to provide a full treatment of capture--recapture theory, but to show concretely why different types of identification errors matter in different ways.

In a simple two-occasion mark--recapture setting, let $n$ denote the number of individuals recorded on the first occasion, $K$ the number recorded on the second occasion, and $k$ the number recorded on both occasions. The estimated population size is then $\widehat{N} = \frac{nK}{k}$.
Intuitively, this estimate uses the fraction of individuals seen again on the second occasion to infer how many individuals were likely present in total.

This simple formula already shows why identification errors have asymmetric ecological consequences. If observations of the same animal are mistakenly split into multiple identities, the apparent number of individuals can increase, and the number of shared identities across occasions can decrease. This tends to inflate the abundance estimate. By contrast, if observations of different animals are mistakenly merged into one identity, the apparent number of individuals can decrease, and the overlap between occasions can be distorted, which tends to reduce the estimate. In extreme cases, if no recaptures remain after misidentification, the estimate becomes undefined.

For a computer scientist, the important point is that these errors are not interchangeable. A method that makes few mistakes overall may still be poorly suited to abundance estimation if it tends to make the wrong kind of mistake. In this setting, it is not enough to ask whether a system is “accurate” on average. We must also ask whether it tends to create false individuals, collapse real individuals together, or distort the re-encounter structure on which ecological inference depends.

Table~\ref{tab:errors} provides simple toy examples of these effects. These examples are not meant to cover all possible cases, but to make explicit how identification mistakes propagate into a standard ecological estimator. This is the broader point developed throughout the paper: methods for automated individual identification should be evaluated not only by overall predictive performance, but by how their errors affect the ecological quantities we ultimately care about.

% Inline duck with guaranteed baseline control
\newcommand{\duckinline}[1]{%
  \raisebox{-0.05\height}{% <-- adjust this number
    \tikz[scale=0.22]{\duck[#1];}%
  }%
}

\definecolor{duckOrange}{HTML}{E69F00}
\definecolor{duckBlue}{HTML}{56B4E9}
\definecolor{duckPurple}{HTML}{CC79A7}
\definecolor{duckTeal}{HTML}{009E73}

\DeclareRobustCommand{\indA}{\duckinline{body=duckOrange!90!white}}
\DeclareRobustCommand{\indB}{\duckinline{body=duckBlue!90!white}}
\DeclareRobustCommand{\indC}{\duckinline{body=duckPurple!90!white}}
\DeclareRobustCommand{\indD}{\duckinline{body=duckTeal!90!white}}

\begin{table}[h]
\centering
\small
\renewcommand{\arraystretch}{1.5}
\setlength{\tabcolsep}{8pt}

\caption{\textbf{Identification errors are not ecologically interchangeable.}
Illustrative cases based on the Lincoln-Petersen estimator show how different kinds of identification errors propagate into abundance estimates in different ways. Here, $n$ and $K$ denote the recorded numbers of individuals on Day~1 and Day~2, $k$ denotes the number of recorded identities shared across days, $\widehat{N}=nK/k$ is the resulting abundance estimate, and $N$ is the true population size. Split errors can create extra apparent individuals and inflate abundance estimates, whereas merge errors can collapse distinct individuals and reduce abundance estimates or make them undefined.}
\begin{tabularx}{\linewidth}{@{}l l l c c c c c l@{}}
\toprule
\textbf{Error} & \textbf{Day 1} & \textbf{Day 2} &
\textbf{$n$} & \textbf{$K$} & \textbf{$k$} & \textbf{$\widehat{N}$} & \textbf{$N$} & \textbf{Bias} \\
\midrule

-- &
\indA\ \indB &
\indA &
2 & 1 & 1 & 2 & 2 & -- \\

\midrule

\textit{Merge} &
\indA &
\indA &
1 & 1 & 1 & 1 & 2 & under \\

\textit{Merge} &
~~~~ \indB &
\indA &
1 & 1 & 0 & \texttt{NaN} & 2 & undefined \\

\midrule

\textit{Split} &
\indA\ \indB &
\indA\ \indB &
2 & 2 & 2 & 2 & 2 & exact \\

\textit{Split} &
\indA\ \indB\ \indC &
\indA\ \indB &
3 & 2 & 2 & 3 & 2 & over \\

\textit{Split} &
\indA\ \indB &
~~~~ \indB\ \indC &
2 & 2 & 1 & 4 & 2 & strong over \\

\bottomrule
\end{tabularx}
\label{tab:errors}
\vspace{-5mm}
\end{table}

\begin{table}[h]
\centering
\renewcommand{\arraystretch}{1.45}
\footnotesize
\caption{Glossary of key ecological, workflow, and machine-learning terms used in the field of individual identification. We clarify terms used throughout the paper and help connect language that is often used differently across ecology, computer vision, and machine learning.}
\begin{tabular}{@{}p{1.65cm} >{\raggedright\arraybackslash}p{2.7cm} p{8.25cm}@{}}
\toprule
\textbf{Term used:} & \textbf{Also known as:} & \textbf{Definition:} \\
\midrule

\textbf{Annotation} & \emph{Manual labeling, label, metadata} &
Information or a process of adding context to a record, including identity, bounding box, viewpoints, or other metadata. \\

\textbf{Database} & \emph{Gallery, catalog, reference set, templates} &
A collection of previously identified individuals and their records, used as the reference against which new observations are compared. \\

\textbf{Detection} & \emph{Sighting, recording, sample, image} &
A record, such as an image or an acoustic recording, that might contain a set of signatures. \\

\textbf{Encounter} & \emph{Occurrence, Event, Observation} &
A set of data, such as a burst of images, a short video, or a sequence of voice calls, collected from the same event or at a short time scale. \\

\textbf{False match} & \emph{False link, merging error} &
An error in which records from different individuals are incorrectly treated as belonging to the same individual. \\

\textbf{Human-in-the-loop} & \emph{Expert-guided workflow} &
A workflow in which automated methods assist with tasks such as filtering or candidate retrieval, while people remain involved in reviewing uncertain or high-impact decisions. \\

\textbf{Identifiable record} & \emph{Usable record, matchable record} &
A record that clearly shows enough of the relevant features to support a reliable identification. \\

\textbf{Mark--recapture} & \emph{Capture--recapture, capture-mark-recapture} &
A family of methods that uses repeated observations of individuals to estimate abundance, survival, movement, or other population parameters. \\

\textbf{Missed link} & \emph{Missed match, splitting error} &
An error in which records from the same individual are incorrectly treated as belonging to different individuals. \\

\textbf{Query} & \emph{New encounter, test encounter} &
A new encounter that is compared against the database to find a match or determine whether it may represent a new individual. \\

\textbf{Signature} & \emph{Morphological characteristic, pattern, vocal/visual signal}  &
A feature or combination of features that helps distinguish one individual from another across encounters, such as coat patterns, scars, fin shapes, whisker spots, or call structure. \\

\textbf{Stable identity} & \emph{Persistent identifier} &
A lasting identifier assigned to an individual so that records and later corrections can be tracked consistently over time. \\

\textbf{Traceability} & \emph{Audit trail, decision history} &
Ability to reconstruct how an identity decision was made, what evidence was used, and how those factors may have changed over time. \\

\midrule

\textbf{Animal ~~~~Re-ID} & \emph{Identification of individual animals} &
The process of recognizing the same individual animal across different observations, such as images, video, or audio recordings. \\

\textbf{Candidate match} & \emph{Retrieved candidate} &
A possible identity returned by the system for review, usually as part of a ranked shortlist. \\

\textbf{Classification} & \emph{Categorization} &
Assigning a query directly to one of a fixed set of known individuals. \\

\textbf{Closed-set setting} & \emph{Identification with only known IDs} &
A setting in which every query is assumed to belong to one of the known individuals already present in the database. \\

\textbf{Encounter-aware split} & \emph{Grouped split} &
An evaluation design in which records from the same encounter are kept together, to avoid leakage between training and testing. \\

\textbf{Embedding} & \emph{Feature vector, representation} &
A numerical representation of data that should capture features useful for distinguishing individuals. \\

\textbf{Evaluation metric} & \emph{Performance metric} &
A numerical measure used to summarize performance, such as identification accuracy or the quality of ranked retrieval results. \\

\textbf{Metric learning} & Representation or similarity learning  &
A machine-learning approach that trains a method to place records from the same individual close together, and records from different individuals farther apart, in an embedding space. \\

\textbf{Object Detection} & \emph{Localization} &
The step of locating an animal, or a relevant body part, within an image, video frame, or sound recording. \\

\textbf{Open-set setting} & \emph{Identification with some unknowns} &
A setting in which some queries may belong to previously unseen individuals, so the system must both recognize known individuals and allow for new ones. \\

\textbf{Retrieval} & \emph{Similarity search, ranking} &
Returning a ranked list of the most similar records in the database for a given query. \\

\textbf{Time-aware split} & \emph{Temporal split} &
An evaluation design in which training uses earlier observations and testing uses later ones, to better reflect real monitoring conditions. \\

\textbf{Verification} & \emph{re-identification, 1-to-1 matching} & A binary classification task that determines whether two observations correspond to the same individual. \\

\bottomrule
\end{tabular}
\label{tab:glossary}
\end{table}

\clearpage

\bibliography{bibliography}
\end{document}